\documentclass[a4paper]{article}
\usepackage{a4wide}
\usepackage{amsmath,amssymb,amsfonts,amsthm,latexsym}
\usepackage{color}
\usepackage{graphicx}
\usepackage{fancyhdr}
\fancyhfoffset{30pt}
\usepackage[pdftex,pdftitle={Voronoi and finite-size particles},
breaklinks=true,bookmarks=true,colorlinks=true,linkcolor=blue,citecolor=blue,allcolors=blue]{hyperref} 
\usepackage{algorithm}
\usepackage{algpseudocode}
\usepackage[sort&compress,authoryear,round]{natbib}
\usepackage{bibentry}
\nobibliography*
\usepackage{doi}
\usepackage{relsize}
\newcommand{\ben} {\begin{equation}}
\newcommand{\een} {\end{equation}}
\newcommand{\be} [1] {\begin{equation} \label{#1}}
\newcommand{\ee} {\end{equation}}
\newcommand{\bse} [1] {\begin{subequations} \label{#1}}
\newcommand{\ese} {\end{subequations}}
\newcommand{\ban} {\begin{eqnarray*} }
\newcommand{\ean} {\end{eqnarray*} }
\newcommand{\bea} {\begin{eqnarray}}
\newcommand{\eea} {\end{eqnarray}}

\def\solidthick{\protect\rule[2pt]{10.pt}{1pt}}

\usepackage{tikz}
\usepackage{tikz-3dplot}
\usepackage{pgfplots}
\usetikzlibrary{shapes,arrows,positioning,calc,through,backgrounds}
\tikzstyle{every picture}+=[remember picture]
\usetikzlibrary{patterns}
\definecolor{lightgray}{rgb}{0.75,0.75,0.75}
\definecolor{darkgreen}{RGB}{0,102,0}
\definecolor{forestgreen}{rgb}{0.2344,0.6992,0.4414}
\definecolor{niceorange}{rgb}{0.9216,0.5059,0.1059}

\newcommand{\revision}[2]{#2}
\newcommand{\revisions}[1]{}

\newcommand{\revisionTwo}[2]{#2}
\newcommand{\revisionsTwo}[1]{}

\begin{document}
\epstopdfsetup{suffix=} 
\author{
\href{http://www-cfd.ifh.uni-karlsruhe.de/uhlmann/home/report.html}
{Markus Uhlmann}\\[.5cm]
Institute for Hydromechanics\\
Karlsruhe Institute of Technology\\
76131 Karlsruhe, Germany\\
{\tt markus.uhlmann@kit.edu}
}
\title{%
  Vorono\"i tesselation analysis of sets of randomly placed  
  finite-size spheres 
}
\date{\today}
\maketitle
\thispagestyle{fancy}
\begin{abstract}
The purpose of this note is to clarify the effect of the finite size of
spherical particles upon the characteristics of their spatial distribution
through a random Poisson process (RPP).
\revision{}{%
  This information is of special interest when using RPP data as a
  reference for the analysis of the spatial structure of a given
  (non-RPP) particulate system, in which case ignoring finite-size
  effects upon the former may yield misleading conclusions.} 
We perform Monte Carlo
simulations in triply-periodic spatial domains,
\revision{%
  and then analyze the particle-centered Vorono\"i tesselations.}{%
  and then analyze the particle-centered Vorono\"i tesselations for
  solid volume fractions ranging from $10^{-5}$ to $0.3$. 
}
\revision{%
  We show that the
  standard-deviation of these volumes decreases with the solid
  volume fraction, the deviation from the value of point sets
  being approximately linear.}{%
  We show that the standard-deviation of these volumes decreases with
  the solid volume fraction, the deviation from the value of point
  sets being reasonably approximated by an exponential function.} 
\revision{%
  The domain size for which the random
  assemblies of finite-size particles are generated has a constraining
  effect if the number of particles per realization is chosen too
  small.}{%
  As can be expected, the domain size for which the random
  assemblies of finite-size particles are generated has a constraining
  effect if the number of particles per realization is chosen too
  small.} 
This effect is quantified, and recommendations are given. 
\revision{%
Finally, we have also revisited the case of random point sets (i.e.\
the limit of vanishing particle diameter).
  We have found that the
  frequently cited data reported by
  Ferenc \& Neda [Physica A, 385:518–526, 2007] 
  is less accurate
  than the earlier data by 
  Tanemura [Forma, 18(4):221–247, 2003].}{%
  We have also revisited the case of random point sets (i.e.\
  the limit of vanishing particle diameter), 
  for which we have confirmed the accuracy of the earlier data by 
  Tanemura [Forma, 18(4):221–247, 2003]. 
}
\end{abstract}
\section{Introduction}
One method to characterize the spatial distribution of particles is
with the help of Vorono\"i tesselation. As initially proposed in the
context of particulate multi-phase flow by \cite{monchaux:10b} and
further elaborated by \cite{monchaux:12}, such a tesselaton can give
valuable information on the tendency of particles to cluster.
Apart from partitioning space and therefore allowing for an
interpretation as the inverse of a local concentration, the
tesselation further generates data on the connectivity
(``neighborhood'') between particles, and it can therefore be directly
used for the purpose of cluster identification and for quantifying
Lagrangian aspects of clustering.
Another benefit is the
intrinsic definition of a clustering threshold \citep{monchaux:10b}
without the need for a priori selection of a length-scale. A 
further important advantage is the availability of a fast algorithm for
performing the tesselation 
(in the present note we either use the Matlab implementation of the
``QHULL'' library, \citealt{barber:96}, or the ``VORO++'' library, 
\citealt{rycroft:09}, both of which provide a scaling of the execution 
time which is linear in the number of particles).  
For these reasons, various analysis methods based upon Vorono\"i
tesselation of particle positions have found relatively wide-spread
use in the particulate flow community
\citep{obligado:11,villalba:12,fiabane:12,tagawa:12,dejoan:13,kidanemariam:13,uhlmann:14a,sumbekova:16,uhlmann:16a,monchaux:17,chouippe:18a}.

The usual procedure in most of these afore-mentioned studies consists
in comparing the statistics of the actual particle distributions in
the respecitve multiphase flow systems 
(either obtained from experimental measurements or from numerical
simulations)
with data for particle distributions obtained by a random Poisson
process (henceforth denoted as ``RPP''), i.e.\ drawing them randomly
from a distribution which is uniform in space.
Statistically significant differences are then a sign of ``structure''
in the particle set, and these features can subsequently be
interpreted on physical grounds.

The spatial distribution of a set of points through an RPP has been
investigated in the literature
\citep[][and references therein]{tanemura:03,ferenc:07}. Although no
analytical results are available, these authors have proposed
empirical fits for the probability distribution of the Vorono\"i cell
volumes as well as providing reference values for the first few
moments which are widely used. 
In some applications the shape of the probability density function
(pdf) of Vorono\"i cell volumes does not change much, 
which means that the data is reasonably described by a single
parameter, i.e.\ the second moment. 
It is then common practice to use the difference between
the measured standard deviation and the literature value of an
RPP-generated set of points as a proxy for the tendency to cluster
\citep{monchaux:10b}.  

Due to the finite size of real-world particles, theoretical results
derived for the spatial distribution of points \citep[such as those
given by][]{ferenc:07} do no longer strictly apply. 
This is due to the fact that there exists a lower bound for the
closest packing of finite-size particles as they cannot overlap. This
aspect as well as the impact of a finite domain size (both in
numerical simulations and in laboratory-experimental measurements) has
so far not been systematically documented.
\cite{oger:96} have observed that the probability distribution of
Vorono\"i cell volumes changes from a Gamma distribution to a Gaussian
when going from one extreme in terms of the solid volume fraction to
the other (i.e.\ from a point set to a close sphere packing). 

Incidentally, let us note that \cite{tagawa:12} have reported on the 
effect of the number of particle samples upon the standard deviation
of Vorono\"i cell volumes determined for numerical data-sets of point
particles.  
\cite{monchaux:12b} have analyzed biases in Vorono\"i-based analysis
of experimental data, focusing on: (i) the effect of projecting 
onto two spatial dimensions from measurements with a laser sheet with
non-vanishing thickness; (ii) sub-sampling due to ``missed''
particles; (iii) slight polydispersity.

In the present note we first revisit the case of point sets, before
addressing the following questions pertaining to the statistics of
spherically-shaped particles distributed through a random Poisson
process:  
%
%
what is the effect of a non-vanishing particle size? 
What is the effect of a finite solid volume fraction?
What is the influence of the relative domain size?
%
%
%
\section{Revisiting the results for sets of points}
We consider the case of $N_p$ points distributed inside of a cubical
domain of side-length $L$. For the sake of generality, we consider
$N_s$ independently drawn such distributions (i.e.\ snapshots). Since
there is no second length-scale, the problem is then completely
determined by the two non-dimensional numbers $N_p$ and $N_s$.

The Vorono\"i tesselation is performed by assuming periodicity of the
field over all three spatial directions. This can be (naively)
implemented algorithmically by (redundant) periodic extension of the
fields and later eliminating the duplicate cells on the borders of the 
fundamental domain.

\begin{figure}
  \begin{minipage}{1.5ex}
    \rotatebox{90}{pdf}
  \end{minipage}
  \begin{minipage}{.45\linewidth}
    \centerline{$(a)$}
    \includegraphics[width=\linewidth]
    {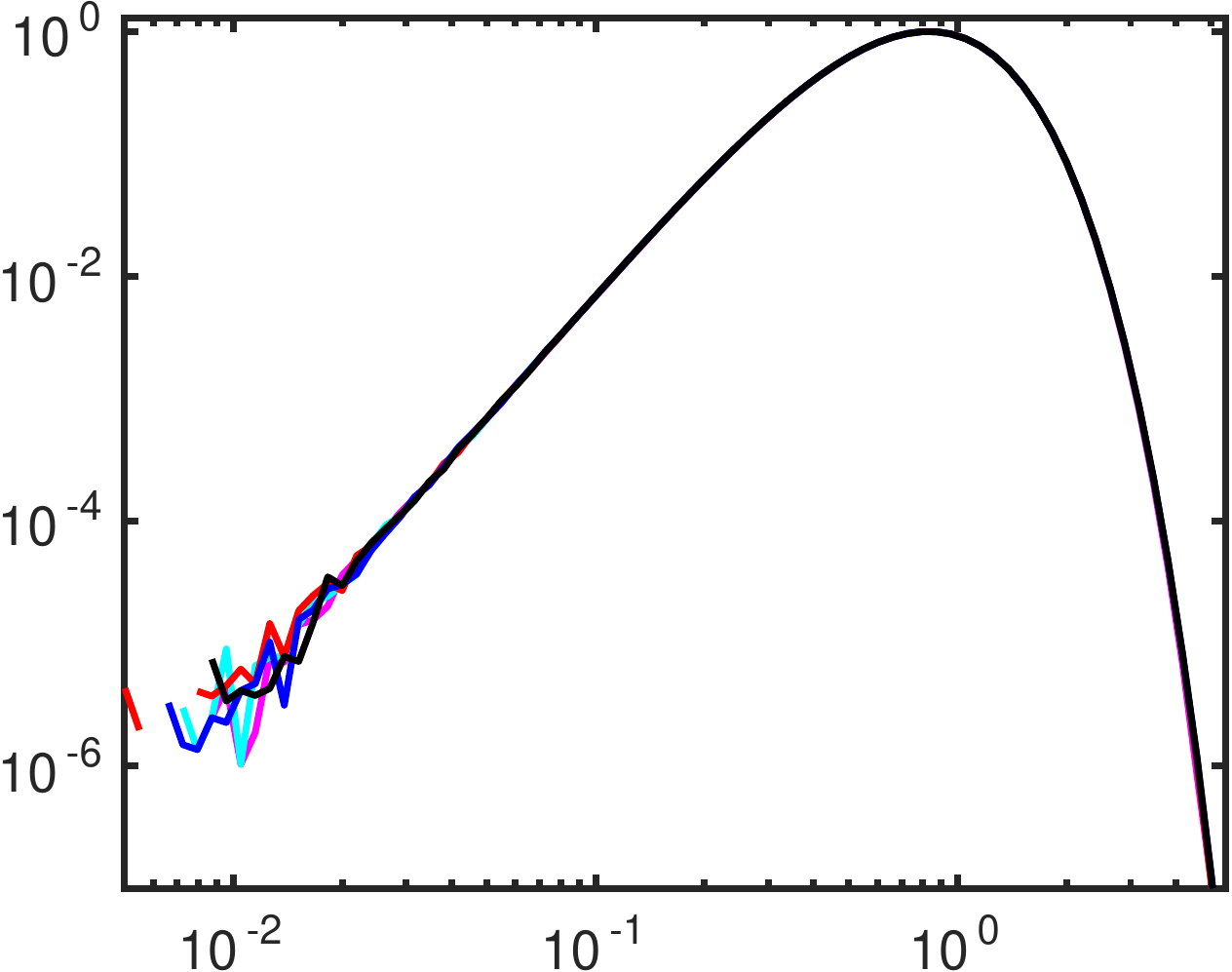}
    \\
    \centerline{$V_{vor}/\langle V\rangle$}
  \end{minipage}
  \hfill
  \begin{minipage}{1.5ex}
    \rotatebox{90}{pdf}
  \end{minipage}
  \begin{minipage}{.45\linewidth}
    \centerline{$(b)$}
    \includegraphics[width=.98\linewidth]
    {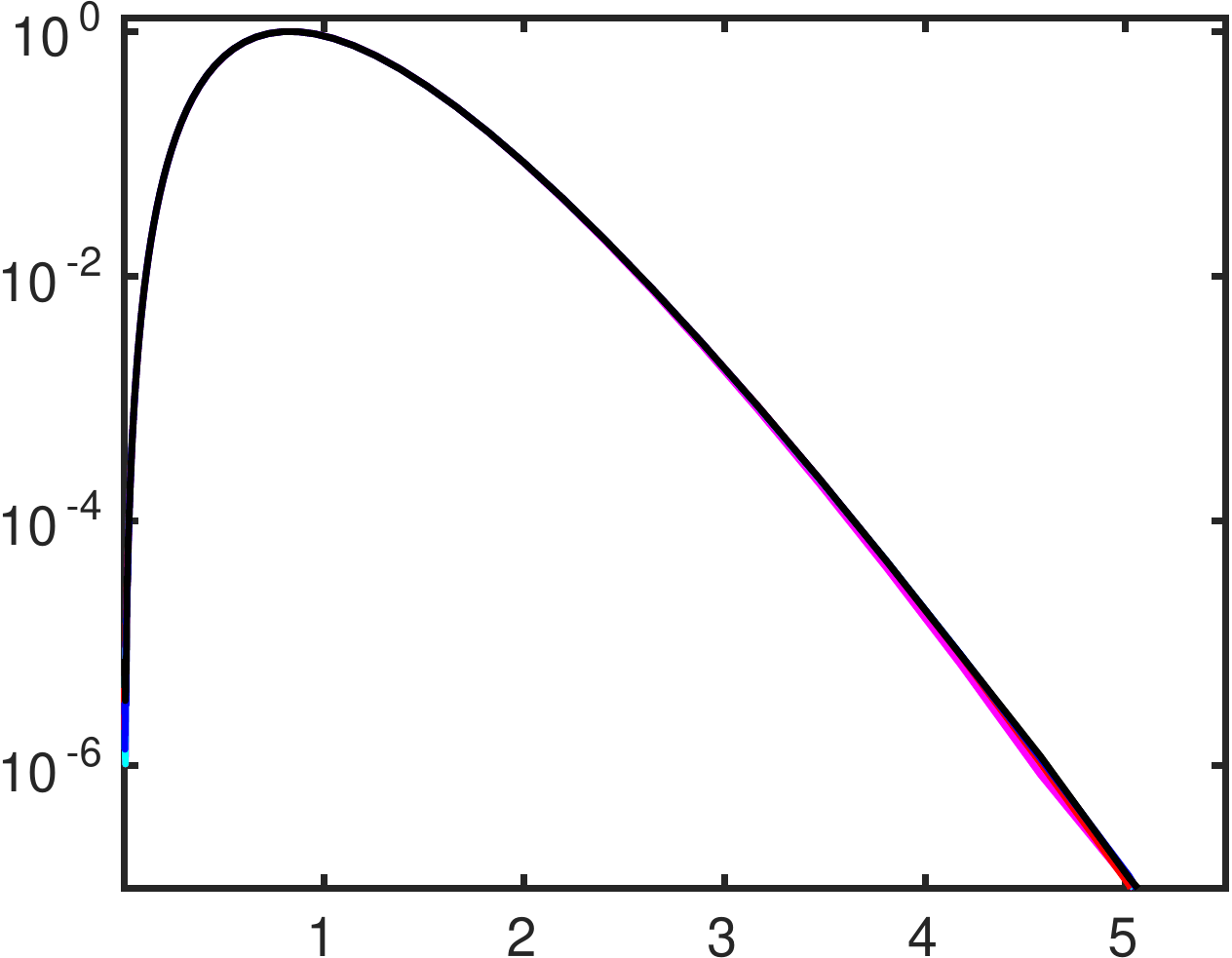}
    \hspace*{-.85\linewidth}\raisebox{.25\linewidth}{
      \begin{minipage}{.45\linewidth}
        \includegraphics[width=\linewidth]
        {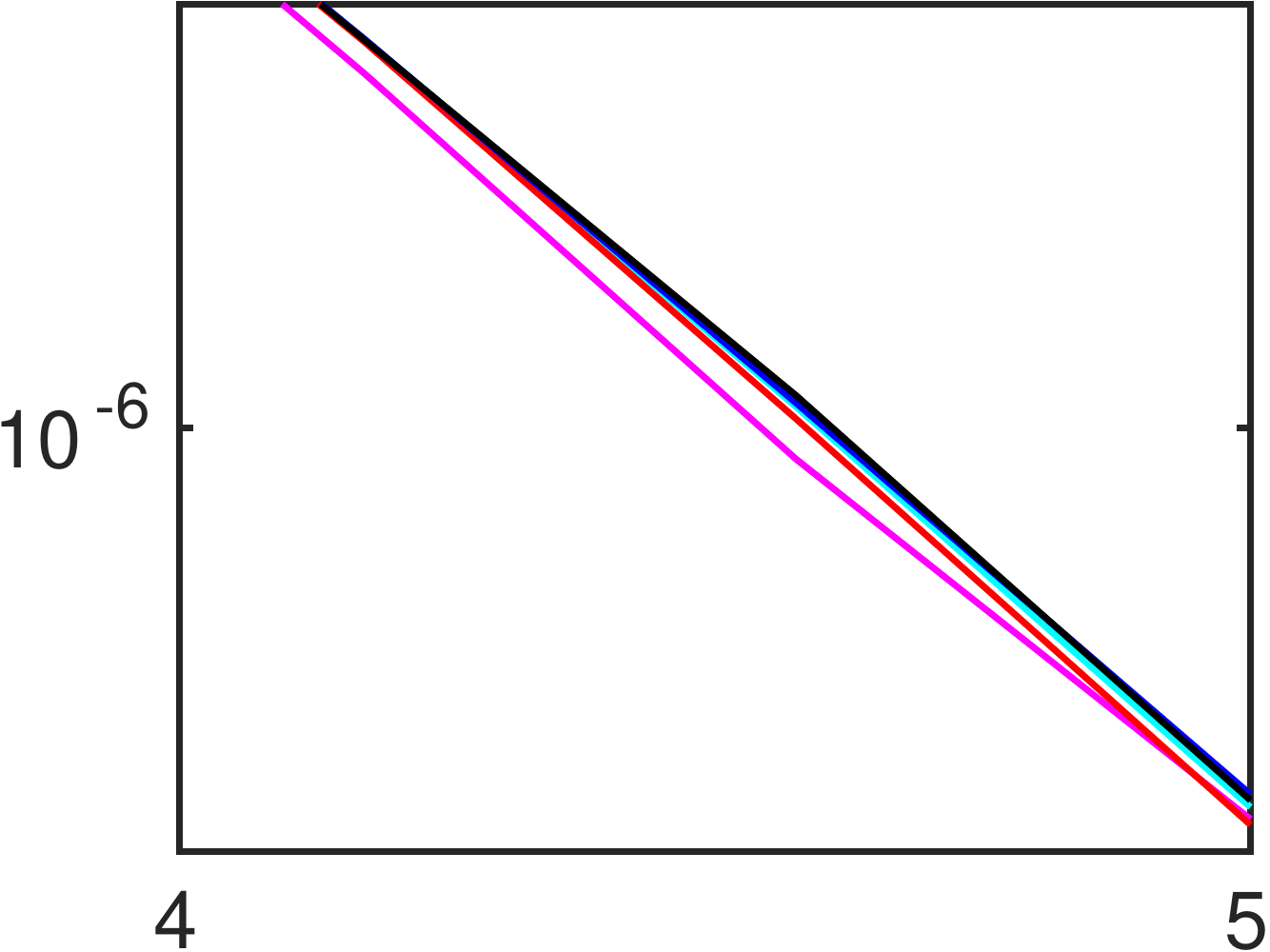}
      \end{minipage}
    }
    \\
    \centerline{$V_{vor}/\langle V\rangle$}
  \end{minipage}
  \caption{%
    Probability density function of the volume of Vorono\"i cells for
    sets of points distributed according to an RPP.
    The number of points per snapshot is indicated by the colors as
    follows:
    {\color{black}\solidthick}, $N_p=10^7$;
    {\color{blue}\solidthick}, $N_p=10^6$;
    {\color{red}\solidthick}, $N_p=10^5$;
    {\color{cyan}\solidthick}, $N_p=10^4$;
    {\color{magenta}\solidthick}, $N_p=10^3$;
    The total number of samples is kept constant
    ($N_{tot}=N_p\,N_s=5\cdot10^9$) in the entire series by adjusting the
    number of snapshots $N_s$ accordingly.  
    Graph $(a)$ shows double logarithmic scaling;
    $(b)$ is the same data in semi-logarithmic scaling.
    The inset shows a zoom of the right tail. 
  }\label{fig-points-vorvol-pdf-1}
\end{figure}
\begin{figure}
  \centering
  \begin{minipage}{2.5ex}
    \rotatebox{90}{$\sigma(V_{vor}/\langle V\rangle)$}
  \end{minipage}
  \begin{minipage}{.45\linewidth}
    \includegraphics[width=\linewidth]
    {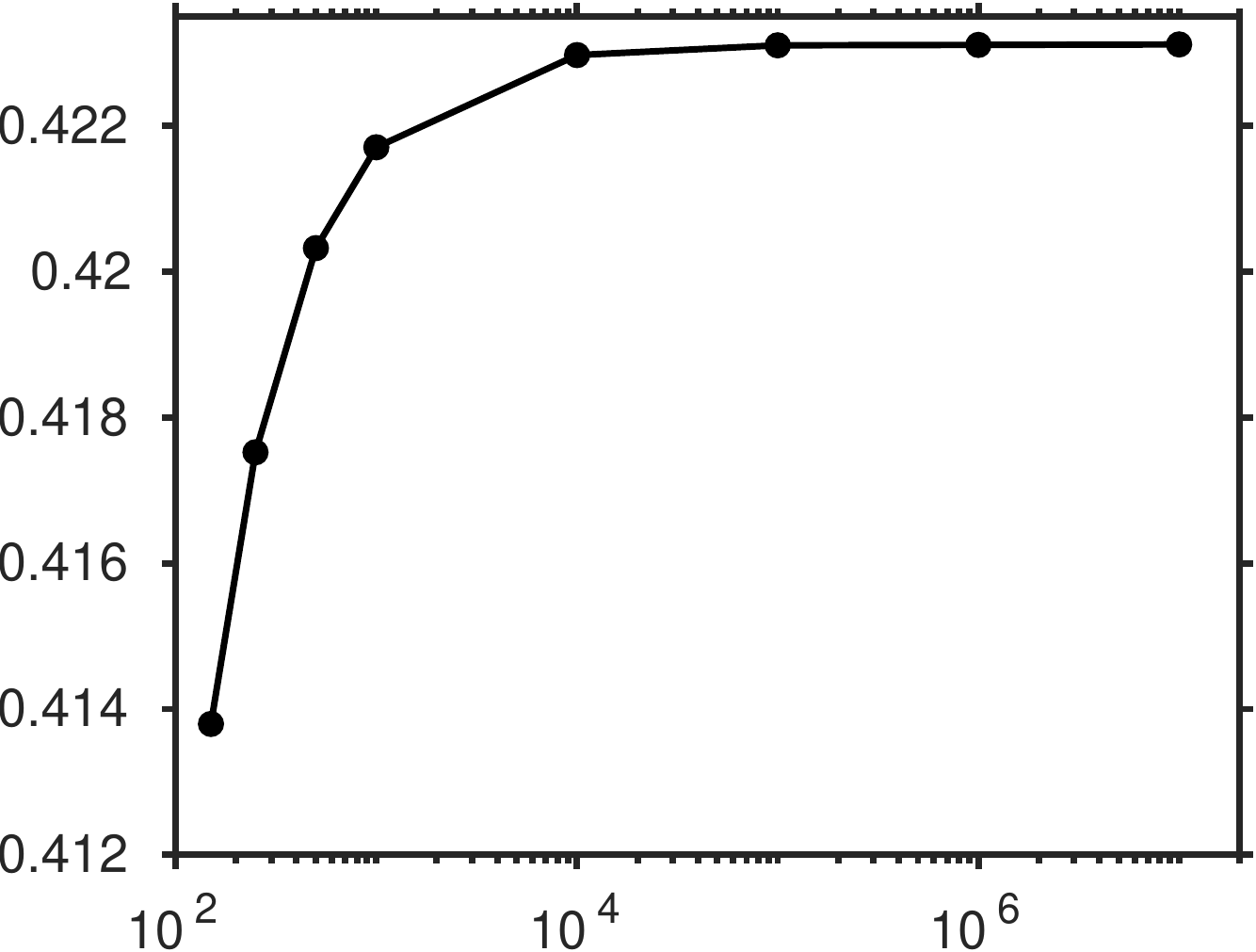}
    \\
    \centerline{$N_p$}
  \end{minipage}
  \caption{%
    Standard deviation of the volume of Vorono\"i cells for
    sets of points distributed according to an RPP, shown as a
    function of the number of particles $N_p$ per snapshot.
    The total number of samples is kept constant
    ($N_{tot}=N_p\,N_s=5\cdot10^9$) in the entire series by adjusting the
    number of snapshots $N_s$ accordingly.
    %
    %
  }\label{fig-points-vorvol-stdev-vs-npart-1}
\end{figure}
\begin{figure}
  \begin{minipage}{2.ex}
    \rotatebox{90}{$\sigma(N_s)$}
  \end{minipage}
  \begin{minipage}{.45\linewidth}
    \centerline{$(a)$}
    \includegraphics[width=\linewidth]
    {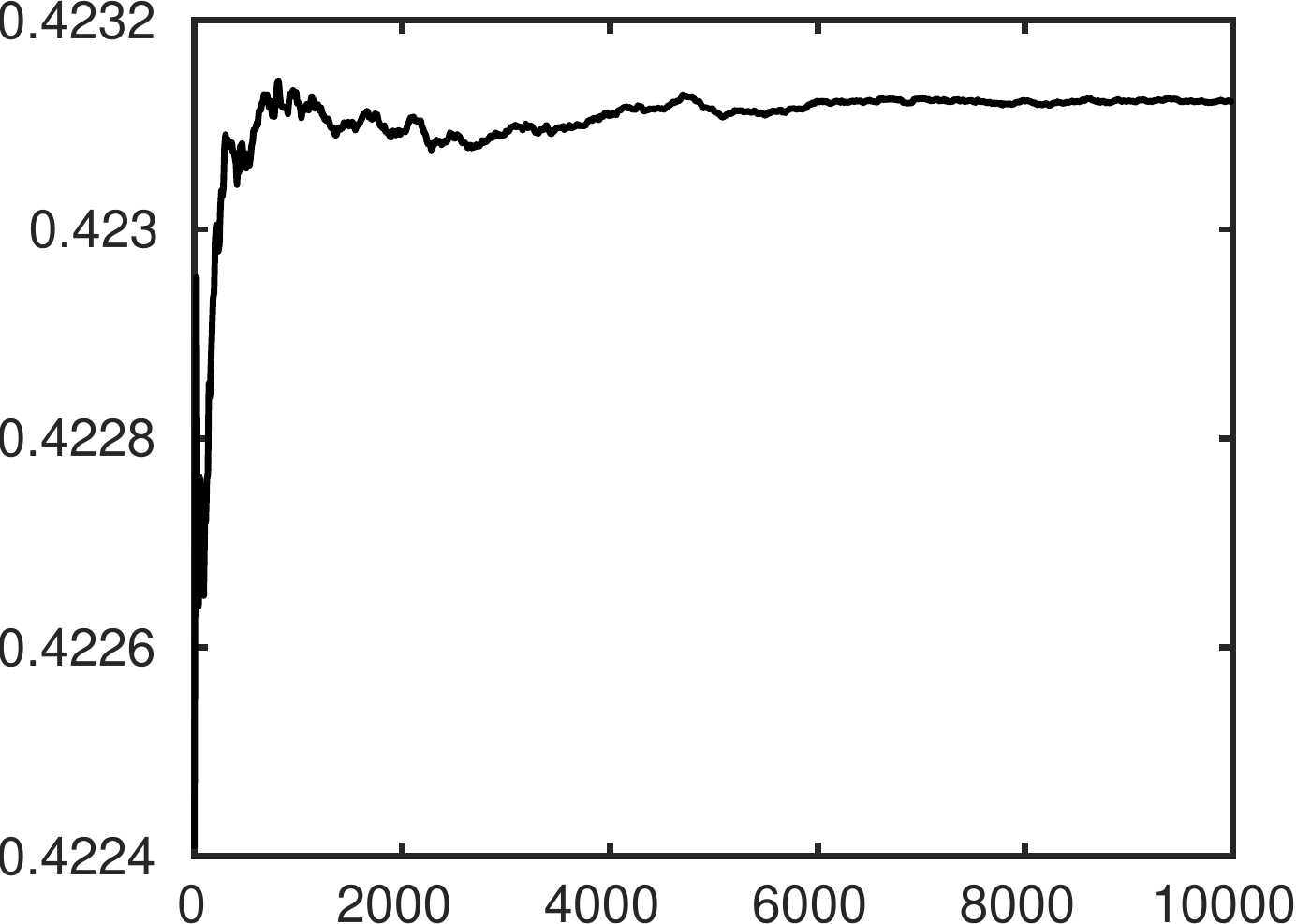}
    \\
    \centerline{$N_s$}
  \end{minipage}
  \hfill
  \begin{minipage}{2.ex}
    \rotatebox{90}{$(\sigma(N_s)-\sigma(10^4))/\sigma(10^4)$}
  \end{minipage}
  \begin{minipage}{.45\linewidth}
    \centerline{$(b)$}
    \includegraphics[width=\linewidth]
    {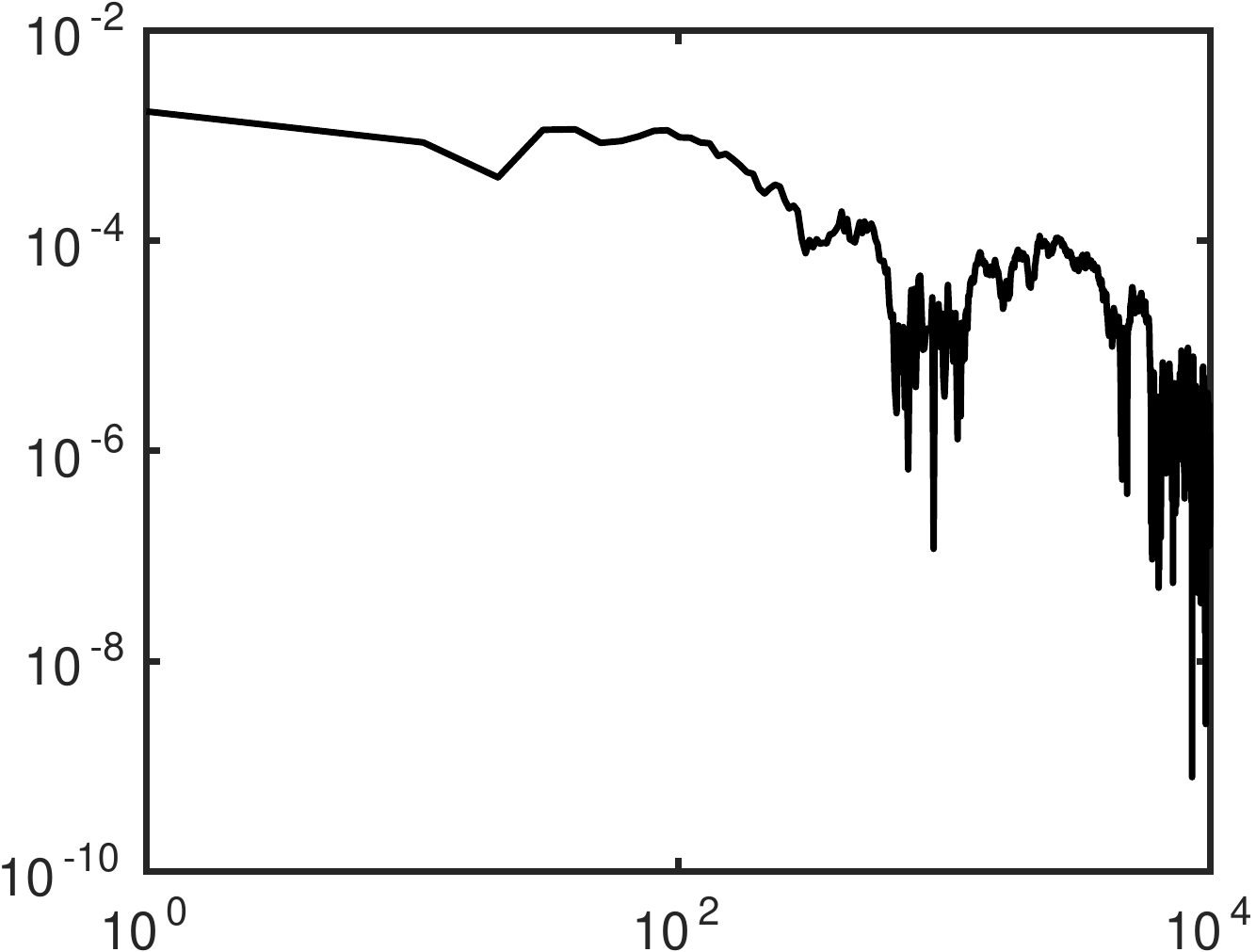}
    \\
    \centerline{$N_s$}
  \end{minipage}
  \caption{%
    Convergence of the standard deviation of the volume of Vorono\"i
    cells for sets of points distributed according to an RPP, shown as
    a function of the number of snapshots $N_s$, for $N_p=10^5$.
    $(a)$ running average;
    $(b)$ relative error with respect to the value obtained for the
    entire sequence. 
  }\label{fig-points-vorvol-stdev-convergence-1}
\end{figure}
Figure~\ref{fig-points-vorvol-pdf-1} shows the probibility density
function obtained by numerical experiments with
$N_{tot}=N_p\,N_s=5\cdot10^9$ samples, by varying the number of particles
per snapshot in the range $N_p=10^3\ldots10^7$ and adjusting the
number of snapshots $N_s$ accordingly. It can be seen that the
differences are marginal. The pdf can be fitted to a three-parameter
(generalized) Gamma distribution, viz.
\begin{equation}\label{equ-three-param-gamma-pdf}
  f(x)=
  \frac{c\,b^{a/c}}{\Gamma(a/c)}x^{a-1}\exp(-b\,x^c)
  \,,
\end{equation}
with the following set of coefficients borne out by our data with
$N_p=10^7$ and $N_s=500$:
\begin{equation}\label{equ-fit1}
  a=4.806\,,\quad
  b=4.045\,,\quad
  c=1.168
  \,.
\end{equation}
These values are very close to those determined by
\cite{tanemura:03} who proposes $a=4.798$. $b=4.040$, $c=1.168$;
they are somewhat distinct from those obtained by
\cite{ferenc:07}.\footnote{%
  We believe that there is a typo in \cite{ferenc:07}: on their
  page~524 they list their three-parameter gamma fit with coefficient
  values as $a=3.24174$, $b=3.24269$, $c=1.26861$,
  which yields a distribution that is quite different from what is
  shown in their figure~6. 
}
Figure~\ref{fig-points-vorvol-stdev-vs-npart-1} shows the influence of
the number of points per snapshot upon the standard-deviation of the
Vorono\"i cell volumes. It can be seen that the influence becomes
insignificant for $N_p\geq10^4$ particles. For smaller numbers, it
turns out that the standard-deviation is underpredicted. This is
due to the small deviations in the right tails (for large cell
volumes) which are visible in the insert of
figure~\ref{fig-points-vorvol-pdf-1}$(b)$. The differences are
presumably caused by an under-representation of very large cell volumes
in ensembles with smaller numbers of particles.

Let us briefly mention the convergence of the second moment with the
number of samples
$N_s$. Figure~\ref{fig-points-vorvol-stdev-convergence-1} exemplarily 
shows the data for the case with $N_p=10^5$. Convergence is obviously
not regular (since there is no order in the sequence), and, therefore,
it needs to be monitored in practice. 

As a bottom line, we state our best approximation of the
standard-deviation of Vorono\"i cell volumes for sets of points drawn 
according to an RPP as follows:\footnote{%
\cite{ferenc:07} state a value of $\sigma=0.43589$, while
\cite{tanemura:03} has found $\sigma=0.42286$. 
}
\begin{equation}\label{equ-sigma-vor-points-value}
  \sigma(V_{vor}/\langle V\rangle)=0.42312\,.
\end{equation}
\section{Sets of finite-size particles}
Now let us turn to the case of a distribution of a set of $N_p$
spherical, mono-dispersed particles with diameter $D$, again placed
by means of an RPP in a cubical box of side-length $L$.
Each draw of particle positions is done through a simple Monte Carlo
method \citep{owen_monteCarloBook:13},
performed under the constraint that no two particles should
overlap. This is performed for each set by first drawing 
individual particle positions sequentially from a uniform and
independent probability distribution (as in the case of sets of points
treated above),
testing each new candidate for geometrical overlap with all previously determined
positions in the set, and, if overlap is found, discarding that
candidate, then re-drawing, and so on until no overlap is detected
(cf.\ algorithm~\ref{algo-finiteSize-algorithm-1}).
\revisionTwo{}{%
  The present procedure is equivalent to random sequential
  addition (RSA) of hard spheres, as sometimes used in the context of
  physical chemistry in order to obtain random sphere arrangements
  \citep{widom:66}.  
  In particular, it has been shown that RSA leads to a so-called
  jamming limit (i.e.\ an upper bound for the solid volume fraction
  $\Phi_s$ 
  where no additional spheres can be placed without overlapping) for
  $\Phi_s^{(RSA)}\approx0.382$ \citep{talbot:91}.
  Note that the RSA limit is significantly lower than the densest
  possible sphere packing ($\Phi_s^{(densest)}\approx0.74$).
  In the present work we have considered solid volume fractions up to
  $0.3$; for values which are even closer to $\Phi_s^{(RSA)}$ it
  becomes challenging to determine a sufficiently large number $N_s$
  of particle assemblies, since the required number of draws
  $N_{draws}$ in algorithm~\ref{algo-finiteSize-algorithm-1} 
  eventually increases exponentially with $\Phi_s$. 
}
\begin{algorithm}[t]
  \caption{%
    Determines a single snapshot, i.e.\ a spatial distribution of $N_p$
    finite-size particles which do not overlap.
    \revisionTwo{}{The number of required draws is stored in the
      variable $N_{draws}$.}
  }
  \label{algo-finiteSize-algorithm-1}
  \revisionTwo{}{
  \begin{algorithmic}[0]
      \State $N_{draws}=0$
    \For{$i=1\ldots N_p$}  \Comment{loop over particle set}
    \State redraw=true
    \While{redraw}
    \State draw ${X}^{(i)}_\alpha$ from uniform distribution
    $\forall\,\alpha=1,2,3$
      \State $N_{draws}\leftarrow N_{draws}+1$
    \If{$\mathbf{X}^{(i)}$ does not lead to overlap with any particle
      $j=1\ldots i-1$} 
    \State redraw=false
    \EndIf
    \EndWhile
    \EndFor
  \end{algorithmic}
}
\end{algorithm}
\revision{
  \begin{figure}
    \begin{minipage}{2.5ex}
      \rotatebox{90}{$\sigma(V_{vor}/\langle V\rangle)$}
    \end{minipage}
    \begin{minipage}{.45\linewidth}
      \centerline{$(a)$}
      \includegraphics[width=\linewidth]
      {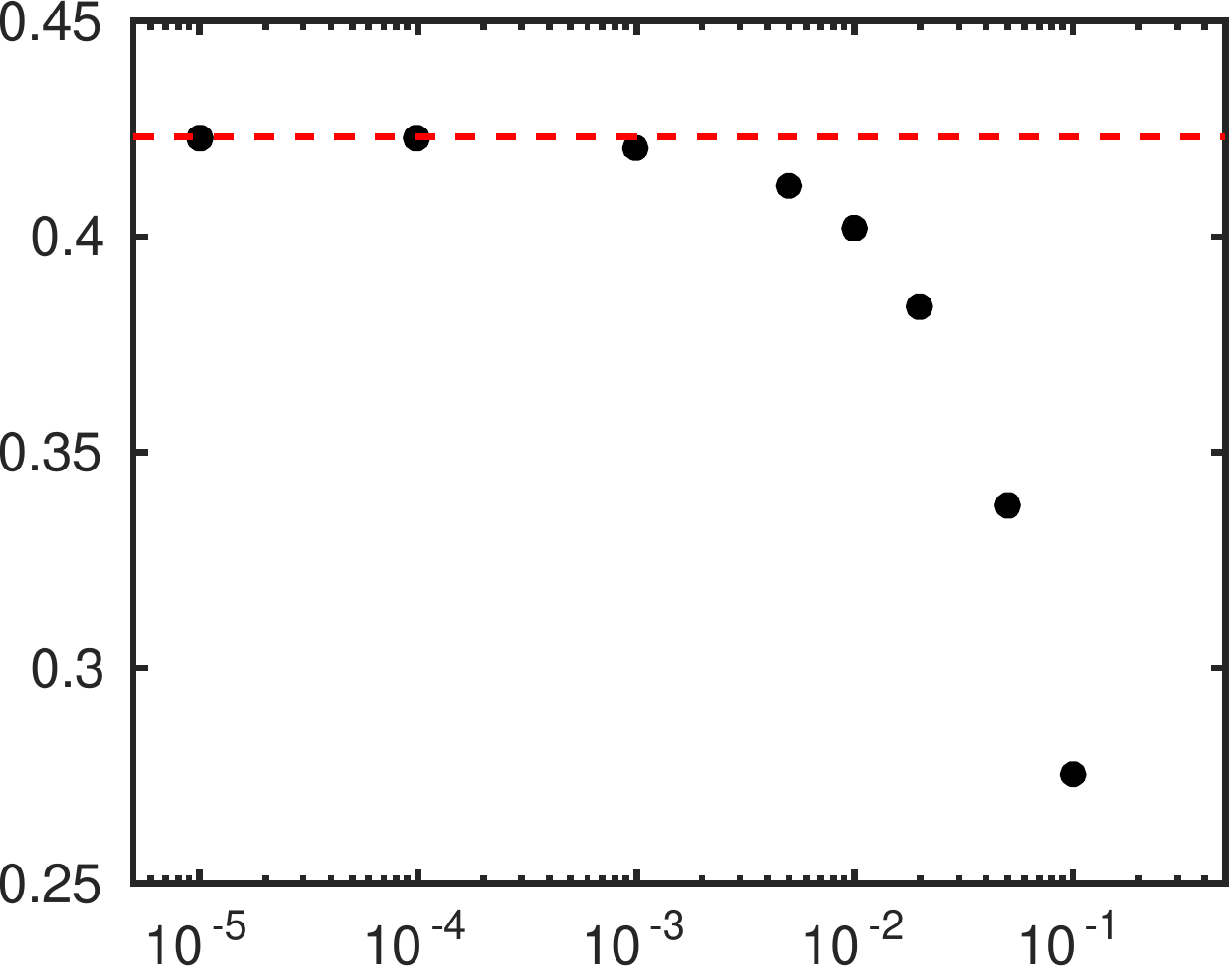}
      \hspace*{-.4\linewidth}\raisebox{.7\linewidth}{%
        {\color{red}$\Phi_s=0$ (points)}} 
      \\
      \centerline{$\Phi_s$}
    \end{minipage}
    \hfill
    \begin{minipage}{2.5ex}
      \rotatebox{90}{$\sigma(0)-\sigma(\Phi_s)$}
    \end{minipage}
    \begin{minipage}{.45\linewidth}
      \centerline{$(b)$}
      \includegraphics[width=\linewidth]
      {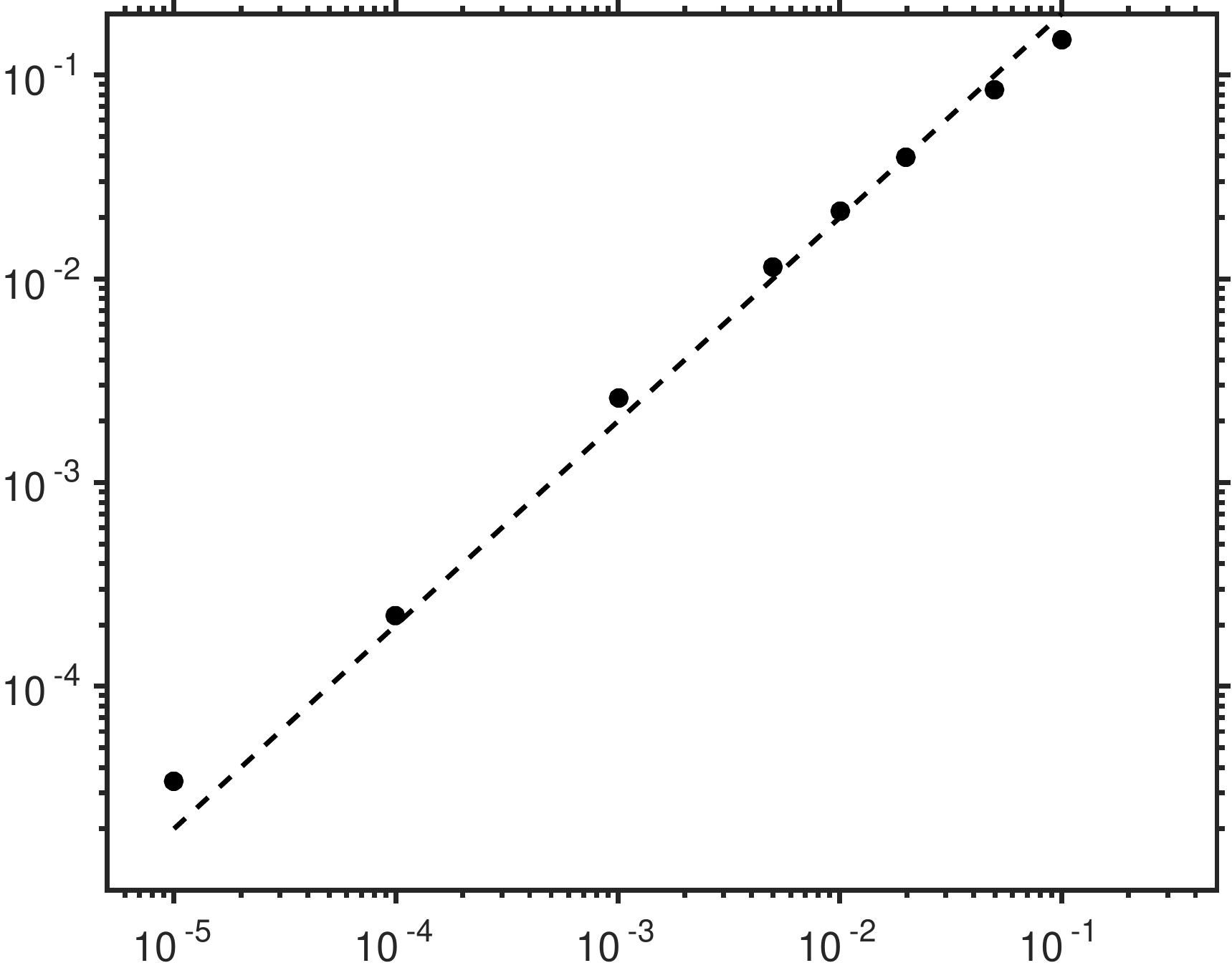}
      \\
      \centerline{$\Phi_s$}
    \end{minipage}
    \caption{%
      Standard deviation of the volume of Vorono\"i cells for
      $N_p=10^5$ finite-size particles distributed according to an RPP,
      as a function of the solid volume fraction $\Phi_s$. The ratio of
      box-size to particle diameter was varied accordingly, i.e.\
      $L/D=(N_p\pi/(6\Phi_s))^{1/3}$. 
      The number of snapshots analyzed in each case was $N_s=2000$, such
      that the total number of samples for each data point measures
      $N_{tot}=2\cdot10^8$.
      The graph in $(a)$ includes the value $\sigma(\Phi_s=0)$ 
      for an RPP distribution of points, as given in
      (\ref{equ-sigma-vor-points-value}).
      The graph in $(b)$ shows the deviation from the point-set value
      in double-logarithmic scale. The dashed line indicates a 
      power-law $\sim\Phi_s$.  
    }\label{fig-finiteSize1-vorvol-stdev-vs-phis-1}
  \end{figure}}{%
  \begin{figure}
    \centering
    \begin{minipage}{2.5ex}
      \rotatebox{90}{$\sigma(V_{vor}/\langle V\rangle)$}
    \end{minipage}
    \begin{minipage}{.45\linewidth}
      \includegraphics[width=\linewidth]
      {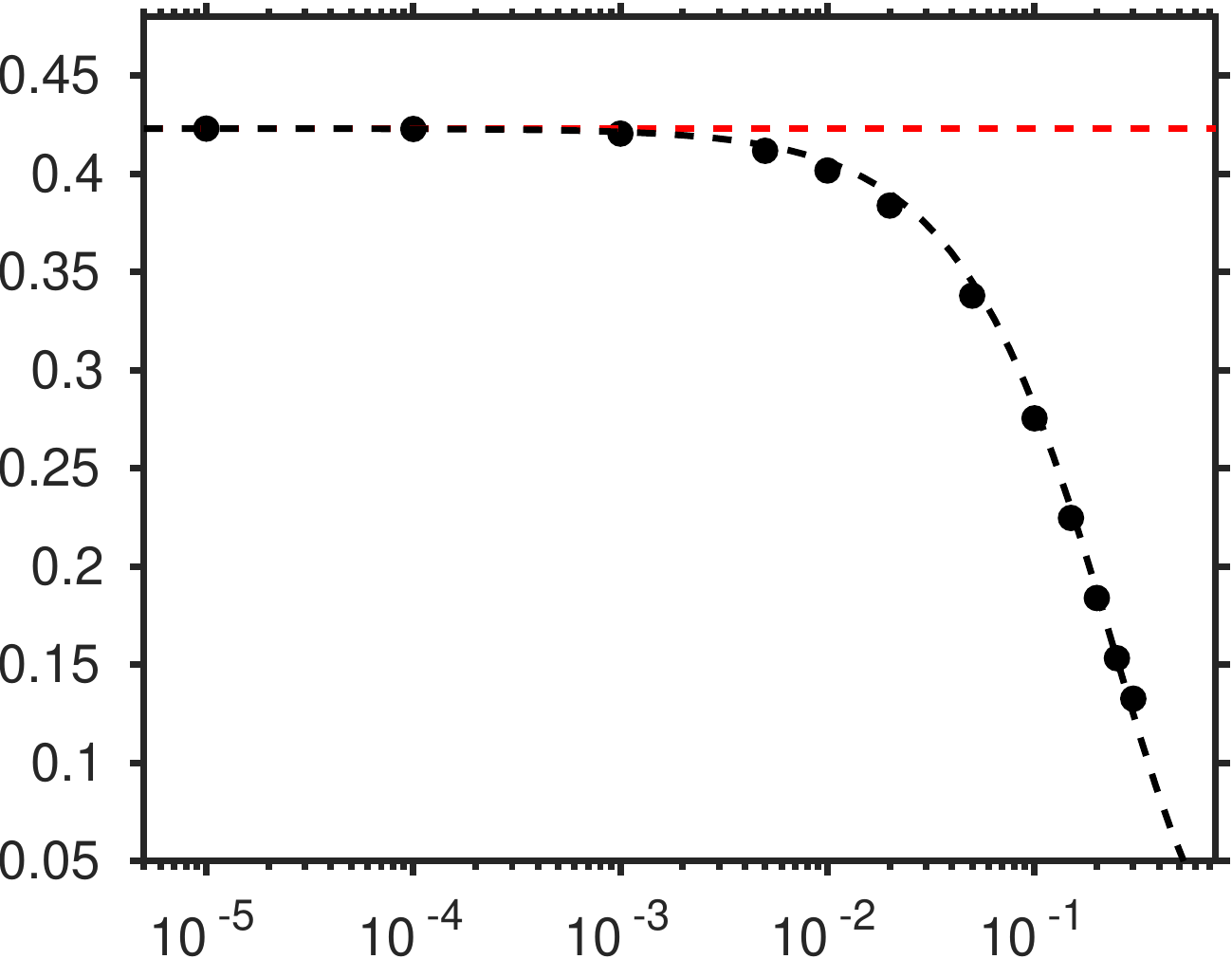}
      \hspace*{-.4\linewidth}\raisebox{.7\linewidth}{%
        {\color{red}$\Phi_s=0$ (points)}} 
      \\
      \centerline{$\Phi_s$}
    \end{minipage}
    \caption{%
      Standard deviation of the volume of Vorono\"i cells for
      $N_p=10^5$ finite-size particles distributed according to an RPP,
      as a function of the solid volume fraction $\Phi_s$. The ratio of
      box-size to particle diameter was varied accordingly, i.e.\
      $L/D=(N_p\pi/(6\Phi_s))^{1/3}$. 
      The number of snapshots analyzed in each case was $N_s=2000$, such
      that the total number of samples for each data point measures
      $N_{tot}=2\cdot10^8$.
      Included is the value $\sigma(\Phi_s=0)$ 
      for an RPP distribution of points, as given in
      (\ref{equ-sigma-vor-points-value}).
      \revision{}{%
        The black dashed line indicates the exponential fit 
        (\ref{equ-best-fit-deviation-solid-vol-frac}).
      }
    }\label{fig-finiteSize1-vorvol-stdev-vs-phis-1}
  \end{figure}}

Here the relevant variables are $N_p$ (number of particles in the
set), $N_s$ (number of particle assemblies analyzed),
$D$ (particle diameter) and $L$ (edge length of the computational
cube), which leads  to similarity under three non-dimensional
parameters. Those can be chosen as: $N_p$, $N_s$ and
$L/D$.
Alternatively, one can use the solid volume fraction 
\begin{equation}\label{equ-def-solid-volume-frac}
  \Phi_s=\frac{\pi}{6}\,N_p\,\left(\frac{D}{L}\right)^3
  \,,
\end{equation}
for the characterisation of a parameter point (instead of either $N_p$
or $L/D$), and/or the total number of samples $N_{tot}=N_pN_s$.
\subsection{Influence of solid volume fraction}
\label{sec-solid-vol-frac}
In a first series we vary the solid volume fraction $\Phi_s$, while
keeping the number of particles per snapshot constant at $N_p=10^5$,
i.e.\ varying the ratio of box-size to particle diameter accordingly,
and systematically using a number of $N_s=2000$ snapshots
($N_{tot}=2\cdot10^8$).
Figure~\ref{fig-finiteSize1-vorvol-stdev-vs-phis-1} shows the
progressive deviation of the standard-deviation of Vorono\"i cell
volumes from the point-set value with increasing solid
volume fraction.
\revision{%
  It can be seen that this deviation is roughly linear
  in $\Phi_s$. The best linear fit to our data is provided by the
  following formula: 
  \begin{equation}\label{equ-best-fit-deviation-solid-vol-frac}
    \sigma(V_{vor}/\langle V\rangle)=
    \sigma(\Phi_s=0)-1.486\,\Phi_s
    \,,
  \end{equation}
  which might be useful as an approximation.}{%
  This deviation can be reasonably fitted with the aid of an
  exponential function in $\Phi_s$, viz.
  %
  \begin{equation}\label{equ-best-fit-deviation-solid-vol-frac}
    \sigma(V_{vor}/\langle V\rangle)=
    \sigma(\Phi_s=0)\cdot\exp\left(
      -4.07\Phi_s
    \right)
    \,,
  \end{equation}
  which might be useful as an approximation.
}
%
\begin{figure}
  \begin{minipage}{1.5ex}
    \rotatebox{90}{pdf}
  \end{minipage}
  \begin{minipage}{.45\linewidth}
    \centerline{$(a)$}
    \includegraphics[width=\linewidth]
    {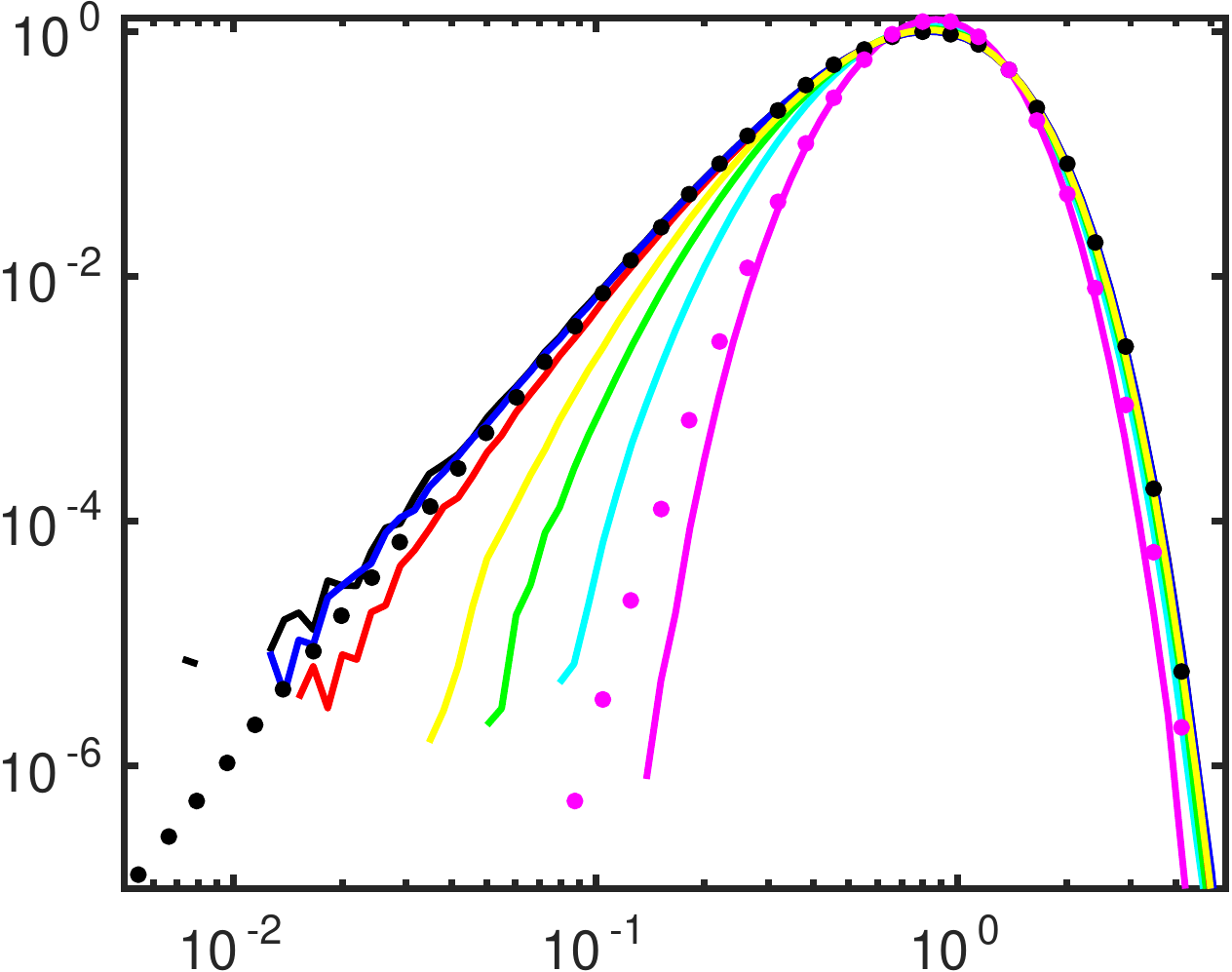}
    \\
    \centerline{$V_{vor}/\langle V\rangle$}
  \end{minipage}
  \hfill
  \begin{minipage}{1.5ex}
    \rotatebox{90}{pdf}
  \end{minipage}
  \begin{minipage}{.45\linewidth}
    \centerline{$(b)$}
    \includegraphics[width=.98\linewidth]
    {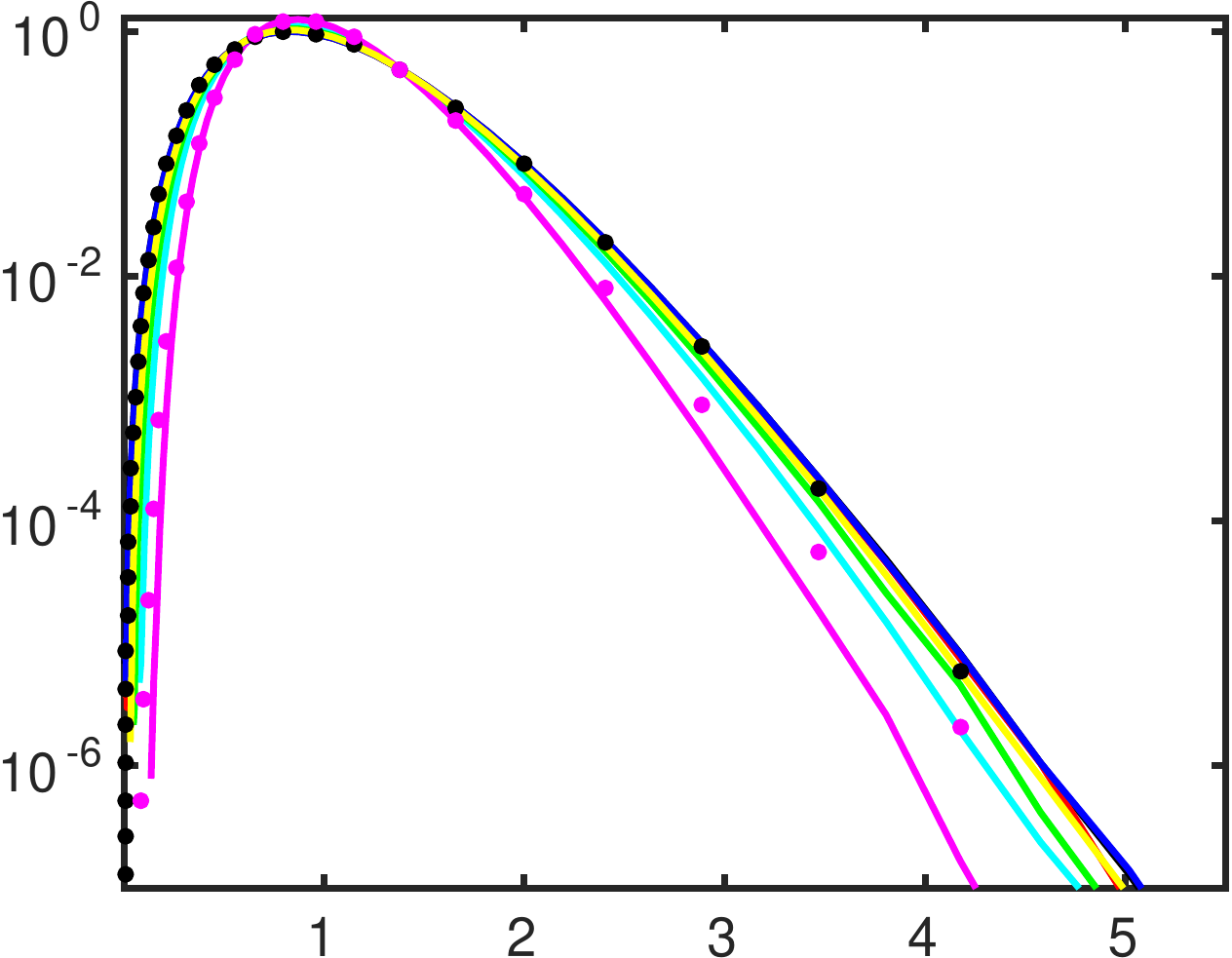}
    \\
    \centerline{$V_{vor}/\langle V\rangle$}
  \end{minipage}
  \caption{%
    Probability density function of the volume of Vorono\"i cells for
    sets of spherical, finite-size particles distributed according to
    an RPP, shown for the data of
    figure~\ref{fig-finiteSize1-vorvol-stdev-vs-phis-1}. 
    The line-styles are as follows: 
    {\color{black}\solidthick},~$\Phi_s=10^{-5}$;
    {\color{blue}\solidthick},~$\Phi_s=10^{-4}$;
    {\color{red}\solidthick},~$\Phi_s=10^{-3}$;
    {\color{yellow}\solidthick},~$\Phi_s=5\cdot10^{-3}$;
    {\color{green}\solidthick},~$\Phi_s=10^{-2}$;
    {\color{cyan}\solidthick},~$\Phi_s=2\cdot10^{-2}$;
    {\color{magenta}\solidthick},~$\Phi_s=5\cdot10^{-2}$.
    Graph $(a)$ shows double logarithmic scaling;
    $(b)$ is the same data in semi-logarithmic scaling.
    The solid circles indicate the respective fits to a generalized
    gamma distribution (\ref{equ-three-param-gamma-pdf}) for the most
    dilute and the densest cases.  
  }\label{fig-finite-size-vorvol-pdf-1}
\end{figure}

The corresponding probability density functions of Vorono\"i cell
volumes for these Monte Carlo experiments are shown in 
figure~\ref{fig-finite-size-vorvol-pdf-1}. 
They can again be fitted reasonably well to three-parameter
(generalized) gamma distributions (\ref{equ-three-param-gamma-pdf}),
with the shape morphing from the 
parameter set ($a=4.81$, $b=4.05$, $c=1.17$) for $\Phi_s=10^{-5}$
(i.e.\ practically identical to the point-set, given in
\ref{equ-fit1})  
to ($a=16.37$, $b=31.73$, $c=0.52$) for $\Phi_s=5\cdot10^{-2}$. 
Note from (\ref{equ-def-solid-volume-frac}) that the solid volume
fraction is equal to the ratio between the volume occupied by a
particle and the mean volume of the Vorono\"i cells, i.e.\
$\Phi_s=V_p/\langle V_{vor}\rangle$, where $V_p=D^3\pi/6$.
Since for non-overlapping particles a Vorono\"i cell cannot be smaller
than the volume occupied by the particle itself, $min(V_{vor})\leq
V_p$, it follows that a lower bound on the normalized Vorono\"i cell
volumes is given by the solid volume fraction itself, viz.\
$min(V_{vor})/\langle V_{vor}\rangle\leq\Phi_s$. 
This lower bound is consistent with what can be observed in
the progressive disappearance of the left tails in
figure~\ref{fig-finite-size-vorvol-pdf-1}$(a)$.
As a consequence of conservation of total volume, the probability of
finding large cells correspondingly decreases, i.e.\ the right tails
are likewise reduced (cf.\
figure~\ref{fig-finite-size-vorvol-pdf-1}$b$).  
%
\begin{figure}
  \begin{minipage}{1.5ex}
    \rotatebox{90}{pdf}
  \end{minipage}
  \begin{minipage}{.45\linewidth}
    \centerline{$(a)$}
    \includegraphics[width=\linewidth]
    {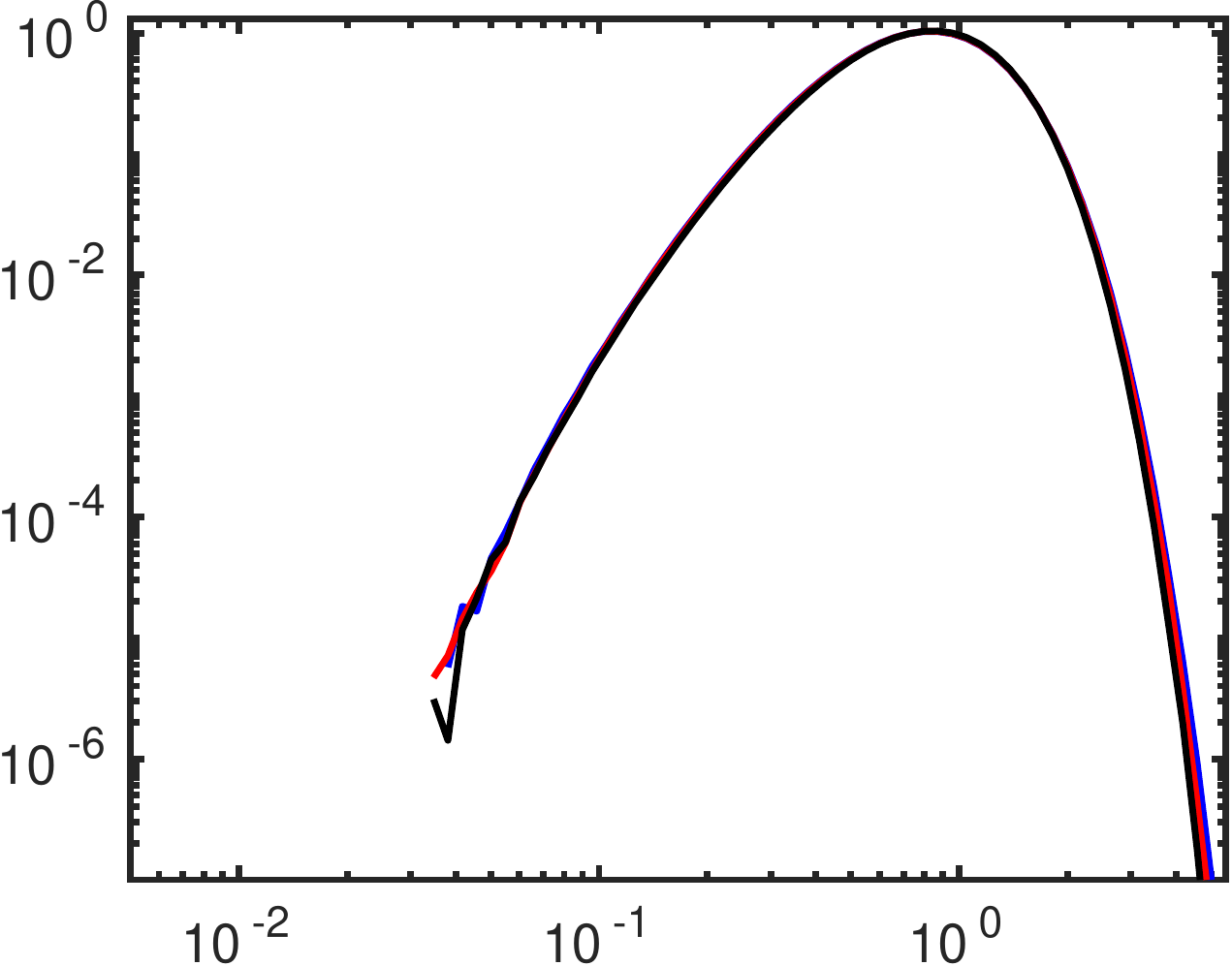}
    \\
    \centerline{$V_{vor}/\langle V\rangle$}
  \end{minipage}
  \hfill
  \begin{minipage}{1.5ex}
    \rotatebox{90}{pdf}
  \end{minipage}
  \begin{minipage}{.45\linewidth}
    \centerline{$(b)$}
    \includegraphics[width=.98\linewidth]
    {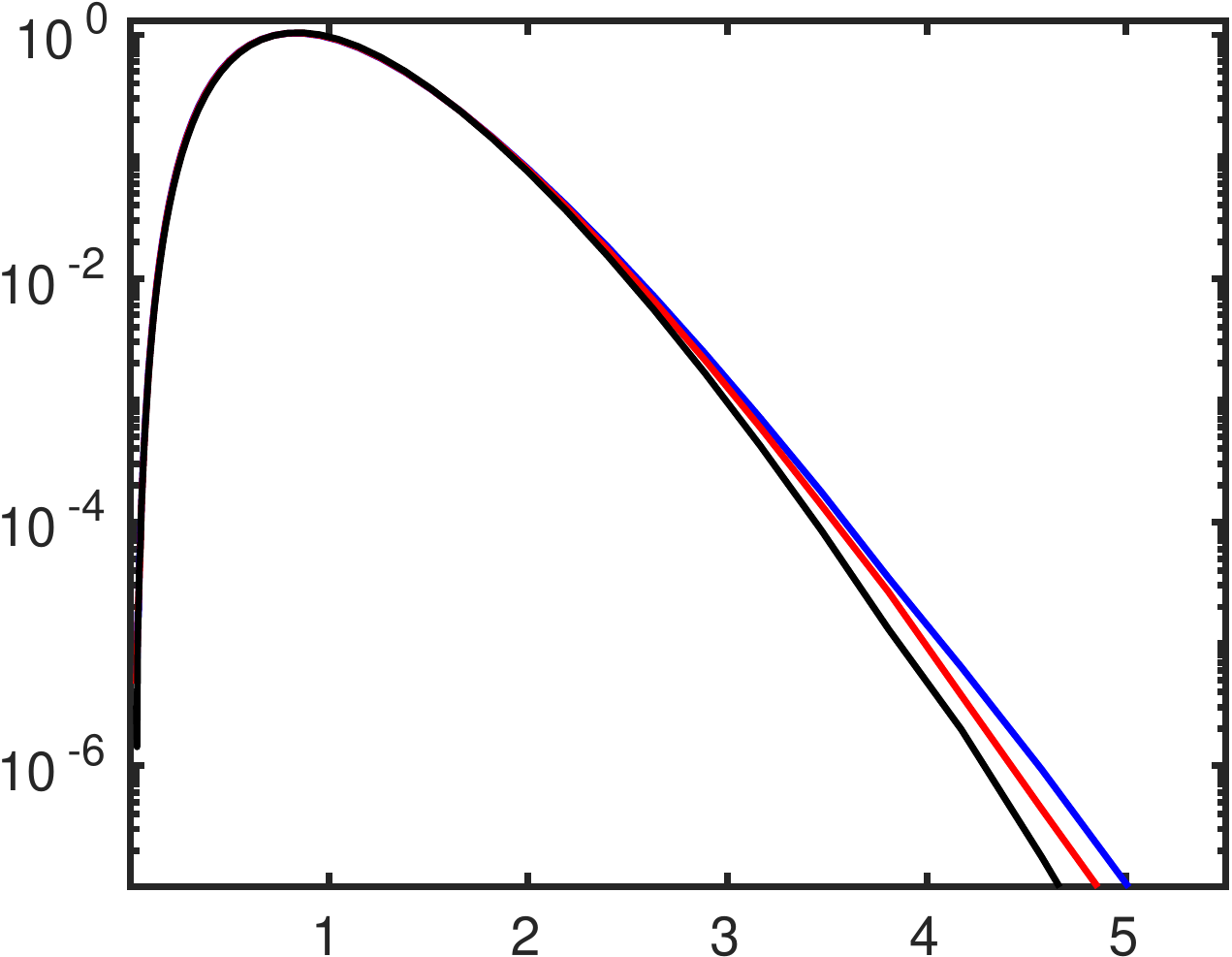}
    \\
    \centerline{$V_{vor}/\langle V\rangle$}
  \end{minipage}
  \caption{%
    Probability density function of the volume of Vorono\"i cells for
    sets of finite-size particles with solid volume fraction
    $\Phi_s=5\cdot10^{-3}$, distributed according to an RPP.
    The relative box-size is indicated by the colors as follows:
    {\color{black}\solidthick}, $L/D=25$;
    {\color{red}\solidthick}, $L/D=35$;
    {\color{blue}\solidthick}, $L/D=150$.
    Graph $(a)$ shows double logarithmic scaling;
    $(b)$ is the same data in semi-logarithmic scaling.
    %
  }\label{fig-finite-size-vorvol-pdf-LoverD-1}
\end{figure}
\subsection{Influence of box size}
Here we first fix a value for the solid volume fraction $\Phi_s$.
We then perform simple Monte Carlo simulations for parameter points
featuring distinct values of the length-scale ratio $L/D$.
This means that for each parameter pair $(\Phi_s,L/D)$ the number of
particles per realization is different, i.e.\
$N_p=\Phi_s(\pi/6)(L/D)^3$. Again we adjust the overall number of
samples per parameter point $(N_{tot}=N_pN_s$) by adapting the number
of snapshots $N_s$; 
for this series we use $N_{tot}\geq2\cdot10^8$. 

\begin{figure}
  \begin{minipage}{2.5ex}
    \rotatebox{90}{$\sigma(V_{vor}/\langle V\rangle)$}
  \end{minipage}
  \begin{minipage}{.45\linewidth}
    \centerline{$(a)$}
    \includegraphics[width=\linewidth]
    {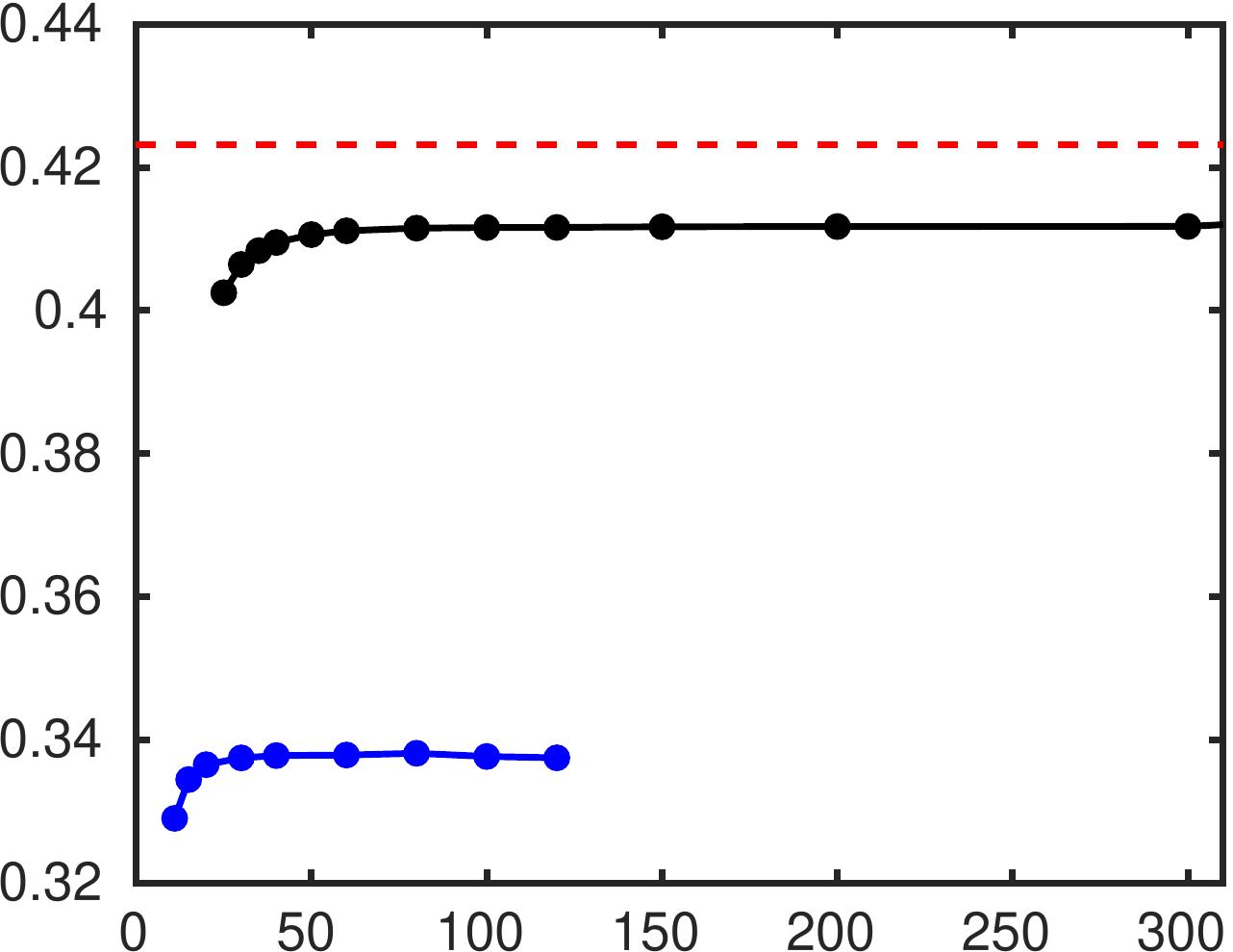}
    \hspace*{-.9\linewidth}\raisebox{.68\linewidth}{%
      {\color{red}$\Phi_s=0$ (points)} 
    }
    \hspace*{+.18\linewidth}\raisebox{.52\linewidth}{%
      {\color{black}$\Phi_s=0.005$}
    }
    \hspace*{-.75\linewidth}\raisebox{.2\linewidth}{%
      {\color{blue}$\Phi_s=0.05$}
    }
    \\
    \centerline{$L/D$}
  \end{minipage}
  \hfill
  \begin{minipage}{2.5ex}
    \rotatebox{90}{$\sigma(V_{vor}/\langle V\rangle)$}
  \end{minipage}
  \begin{minipage}{.45\linewidth}
    \centerline{$(b)$}
    \includegraphics[width=\linewidth]
    {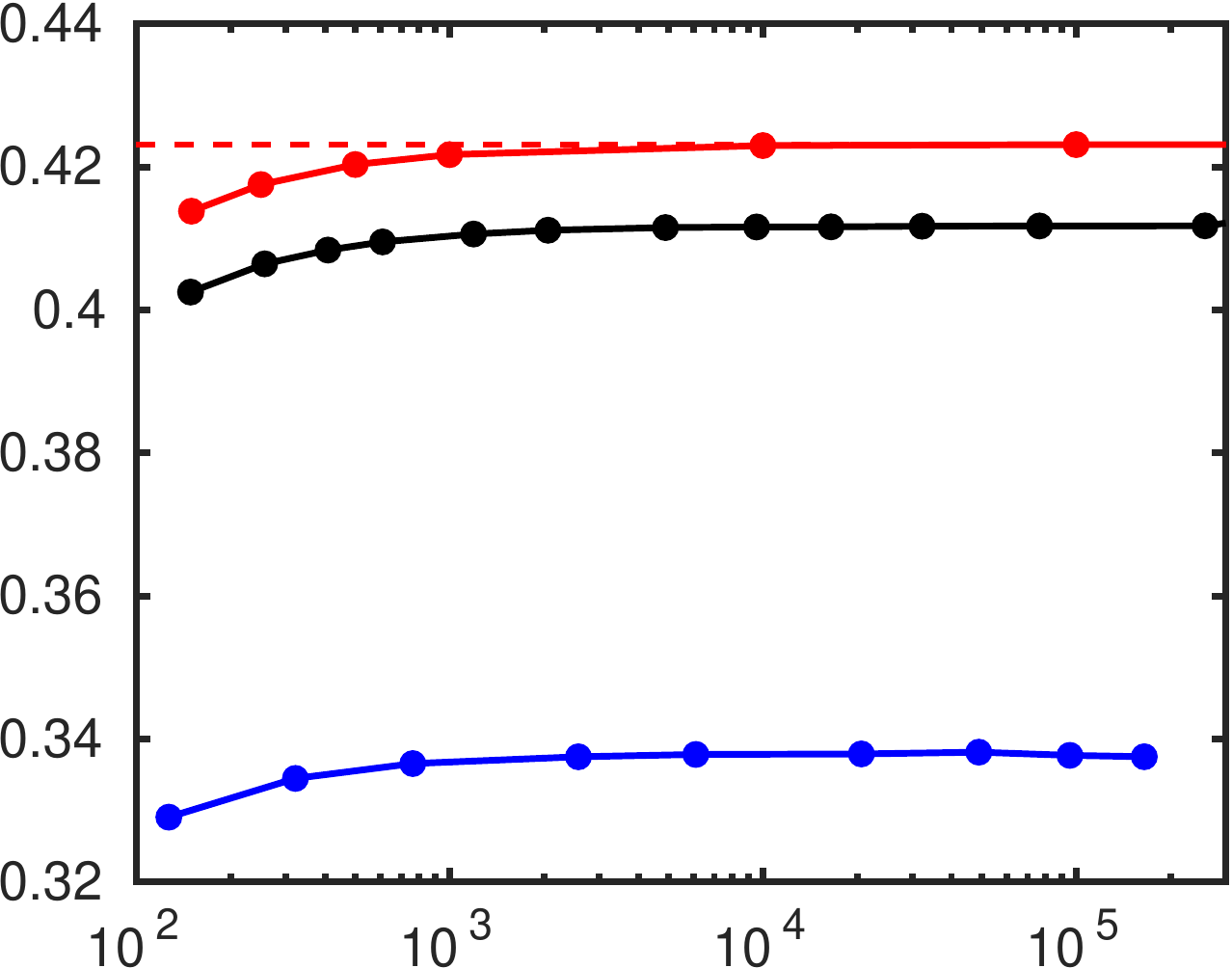}
    \hspace*{-.3\linewidth}\raisebox{.7\linewidth}{%
      {\color{red}$\Phi_s=0$}
    }
    \hspace*{-.4\linewidth}\raisebox{.52\linewidth}{%
      {\color{black}$\Phi_s=0.005$}
    }
    \hspace*{-.25\linewidth}\raisebox{.225\linewidth}{%
      {\color{blue}$\Phi_s=0.05$}
    }
    \\
    \centerline{$N_p$}
  \end{minipage}
  \caption{%
    Standard deviation of the volume of Vorono\"i cells for
    finite-size particles distributed in space according to an RPP,
    shown as: 
    $(a)$ a function of the relative box-size $L/D$; 
    $(b)$ a function of the number of particles per box $N_p$. 
    Each line corresponds to a different value of the solid volume
    fraction, as indicated. 
    %
    %
    The open symbols in blue color in graph $(b)$ correspond to the
    point-particle data of
    figure~\ref{fig-points-vorvol-stdev-vs-npart-1}.  
    %
    %
    %
  }\label{fig-finite-size-vorvol-stdev-vs-npart-1}
\end{figure}

\begin{figure}
  \centering
  \begin{minipage}{2.5ex}
    \rotatebox{90}{$\sigma(V_{vor}/\langle V\rangle)/\sigma(V_{vor}/\langle V\rangle)^\ast$}
  \end{minipage}
  \begin{minipage}{.45\linewidth}
    \includegraphics[width=\linewidth]
    {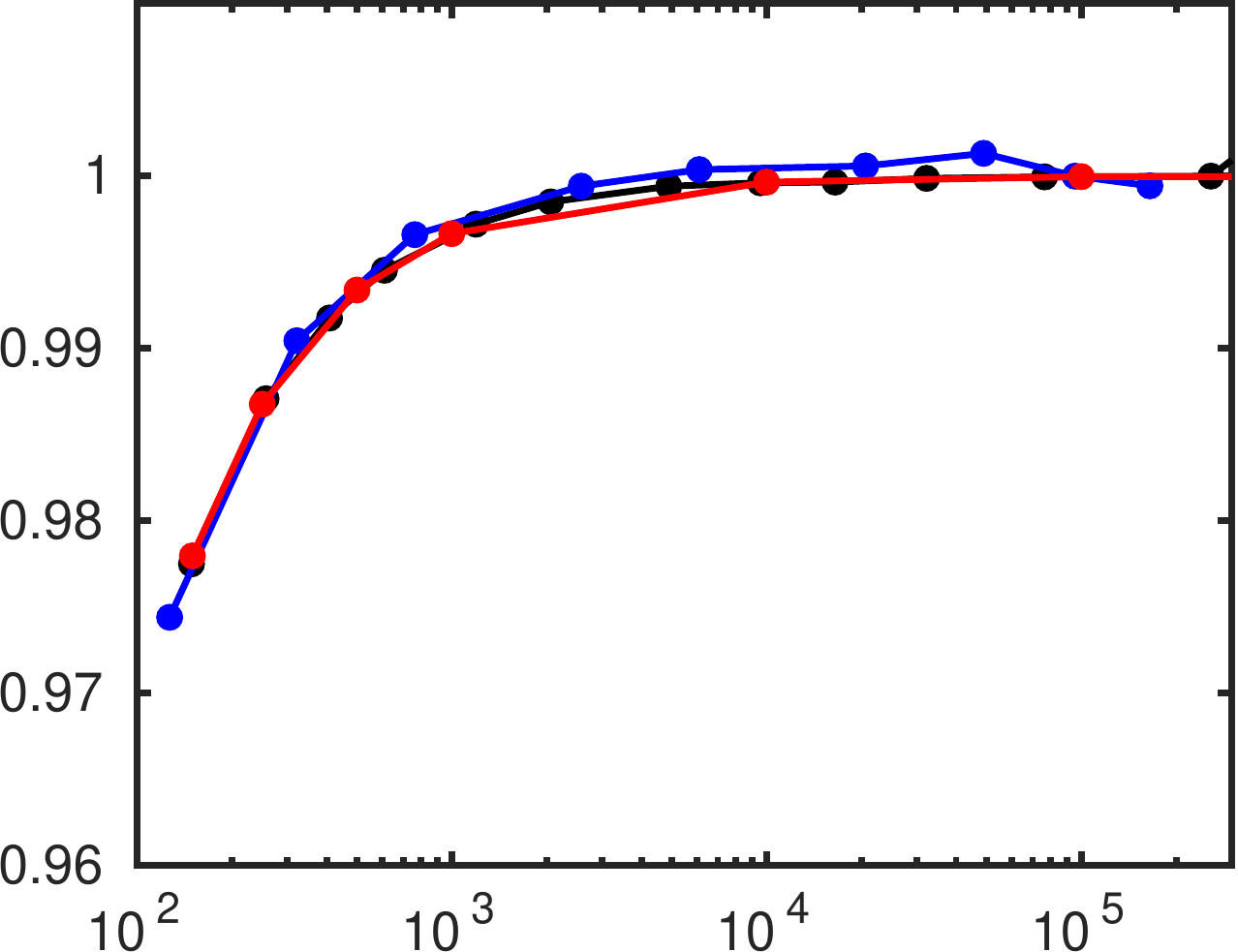}
    \\
    \centerline{$N_p$}
  \end{minipage}
  \caption{%
    The same data as in
    figure~\ref{fig-finite-size-vorvol-stdev-vs-npart-1}$(b)$, but
    the ordinate is normalized with the asymptotic value of
    $\sigma(V_{vor}/\langle V\rangle)$ for large $N_p$. 
    The line-styles are as follows: 
    {\color{red}\solidthick},~$\Phi_s=0$ (points); 
    {\color{black}\solidthick},~$\Phi_s=5\cdot10^{-3}$;
    {\color{blue}\solidthick},~$\Phi_s=5\cdot10^{-2}$.
  }\label{fig-finite-size-vorvol-stdev-vs-npart-normalized-1}
\end{figure}

It can be seen from the pdfs shown in
figure~\ref{fig-finite-size-vorvol-pdf-LoverD-1} that the smallest
box-sizes lead to constraints which are particularly visible in the
far right tails of the distribution (i.e.\ the probability of finding
very large Vorono\"i cells is progressively reduced). 
Figure~\ref{fig-finite-size-vorvol-stdev-vs-npart-1} then shows that
the box-size effect upon the standard deviation is rather mild as
compared to the effect of the solid volume fraction analyzed in
\S~\ref{sec-solid-vol-frac}. 
A significant influence only starts to be noticeable for very small 
relative sizes of the spatial domains $L/D$. It can be seen in the
figure that the value of $L/D$ at which the domain becomes ``small''
depends upon the solid volume fraction.
In figure~\ref{fig-finite-size-vorvol-stdev-vs-npart-normalized-1} we
have normalized the data by its asymptotic value for large domains 
(denoted as $V_{vor}/\langle V\rangle^\ast$). 
As expected, the curves for the two different solid volume fractions
collapse up to statistical uncertainty; they also collapse with
the normalized curve for the point-set. We can then see that ``small''
simply means that the number of particles is below a certain
threshold, say $N_p\leq10^4$.

This threshold can also be argued via a detour, defining a critical
length-scale ratio $(L/\ell)_{crit}$ below which the domain can be
considered as ``small'' 
(here ``$\ell$'' stands for a characteristic length scale of the
random particle arrangement). 
Let us assume that the characteristic length can be defined from the
mean cell volume of a Vorono\"i cell, viz.\ $\ell=\langle
V_{vor}\rangle^{1/3}$.
Then, from the definition $\langle V_{vor}\rangle=L^3/N_p$ it directly
follows that $L/\ell=N_p^{1/3}$, indicating that the critical value
will only depend upon the number of particles, as indeed observed in 
figure~\ref{fig-finite-size-vorvol-stdev-vs-npart-normalized-1}. 
%
\section{Conclusions}
We have investigated random arrangements of point sets and of
non-overlapping, finite-size particles in cubical domains assumed to
be tri-periodic. The analysis was performed with the aid of Vorono\"i
tesselation, a tool which is frequently employed in the context of
particulate flow and other branches of natural sciences and
engineering.
With this study we have addressed several aspects 
that have previously not been documented in a systematic way. 

First, we have revisited the case of point sets. It was shown that
the statistics of a random arrangement (assigned through a random
Poisson process, RPP, which was repeated $N_s$ times per parameter
point) depend upon the number of particles $N_p$, even if the total
number of samples (i.e.\ the product $N_pN_s$) is sufficiently large.
\revision{%
  We have also noticed that the reference data of \cite{ferenc:07} for
  three-dimensional Vorono\"i tesselation is less accurate than earlier
  data proposed by \cite{tanemura:03}.}{}
Our present data leads to the
following best fit to a generalized three-parameter Gamma
distribution (\ref{equ-three-param-gamma-pdf}): 
\begin{equation}\nonumber
  a=4.806\,,\quad
  b=4.045\,,\quad
  c=1.168
  \,,
\end{equation}
which yields as value for the standard deviation:
\begin{equation}\nonumber
  \sigma(V_{vor}/\langle V\rangle)=0.42312\,.
\end{equation}
\revision{}{%
  This value is very close to the result stated by \cite{tanemura:03}, 
  while if differs by roughly 3\% from the result of \cite{ferenc:07}.  
}

Second, we have analyzed spatial arrangements of finite-size,
spherical particles generated through an RPP. The problem depends upon
two non-dimensional parameters (apart from the total number of samples
$N_pN_s$): the solid volume fraction and the relative domain size.
Through numerical evaluation it is demonstrated that the influence of
the solid volume fraction $\Phi_s$ upon the standard deviation of the
Vorono\"i cell volumes can be very significant,
\revision{%
  with the deviation
  from the value for a point-set scaling roughly linearly with
  $\Phi_s$.}{%
  with the deviation from the value for a point-set increasing
  exponentially with $\Phi_s$ over the investigated range of $10^{-5}$
  to $0.3$.}
An approximation which might be useful in practical
applications is provided in
equation~(\ref{equ-best-fit-deviation-solid-vol-frac}).

Concerning the box-size dependence, it is found that -- as expected --
the low-order statistics of the tesselation are constrained if the
box-size is below a certain threshold value.
This threshold is independent of the solid volume fraction if
expressed in terms of the number of particles per realization.
\revision{%
  As a very rough guide, a number $N_p\geq10^4$ should be chosen.}{%
  If a number $N_p\geq10^4$ is chosen, the relative deviation is kept
  below $3\cdot10^{-4}$. 
} 
\section*{Acknowledgements}
I am indebted to two anonymous referees who have made suggestions
which have significantly improved the manuscript. 
The computations were partially performed at SCC Karlsruhe.
The computer resources, technical expertise and assistance
provided by this center are thankfully acknowledged.
%
\bibliographystyle{unsrtnat}
\setlength{\bibsep}{.4ex}
\addcontentsline{toc}{section}{References}

\end{document}